\documentclass[pdftex,twocolumn,epjc3]{svjour3}          % twocolumn

\RequirePackage[T1]{fontenc}

\smartqed  % flush right qed marks, e.g. at end of proof

\RequirePackage{mathptmx}      % use Times fonts if available on your TeX system
\RequirePackage{flushend}
\RequirePackage[numbers,sort&compress]{natbib}
\RequirePackage[colorlinks,citecolor=blue,urlcolor=blue,linkcolor=blue]{hyperref}

\usepackage{latexsym}
\usepackage{amsmath}
\usepackage{amssymb}
\usepackage{latexsym}
\usepackage{amsfonts}
\usepackage{graphicx}
\usepackage{mathrsfs}
\usepackage{hyperref}

\journalname{Eur. Phys. J. C}

\begin{document}

\title{Constraints on scalar-tensor theory of gravity by the recent observational results on gravitational waves}
\author{Yungui Gong\thanksref{e1,addr1}
        \and
       Eleftherios Papantonopoulos\thanksref{e2,addr2}
       \and
       Zhu Yi\thanksref{e3,addr1}
}

\thankstext{e1}{e-mail: yggong@hust.edu.cn}
\thankstext{e2}{e-mail: lpapa@central.ntua.gr}
\thankstext{e3}{e-mail: yizhu92@hust.edu.cn}

\institute{School of Physics, Huazhong University of Science and Technology, Wuhan, Hubei 430074, China \label{addr1}
\and
Department of Physics, National Technical University of Athens, Zografou Campus GR 157 73, Athens, Greece \label{addr2}
}

\date{Received: date / Accepted: date}

\maketitle

\begin{abstract}
The speed of gravitational waves provides us a new tool to test alternative theories of gravity.
The constraint on the speed of gravitational waves from GW170817 and GRB170817A is used to test
some classes of Horndeski theory. In particular, we consider the coupling of a scalar field to Einstein tensor and the coupling of the
Gauss-Bonnet term to a scalar field. The coupling strength of the Gauss-Bonnet coupling is constrained
to be in the order of $10^{-15}$.
In the Horndeski theory we show that in order for this theory to satisfy the stringent constraint on the speed of GWs the
mass scale $M$ introduced in the non-minimally derivative coupling is constrained to be in the range
$10^{15}\text{GeV}\gg M \gtrsim 2\times 10^{-35}$GeV taking also under consideration the early
times upper bound for the mass scale $M$. The large mass ranges require no fine-tuning because the effect
of non-minimally derivative coupling is negligible at late times.
\end{abstract} %end of abstract

%\keywords{inflation, cosmology of theories beyond the SM}
%\preprint{1711.04102}

\section{Introduction}

The detection of gravitational waves (GWs) by the Laser Interferometer Gravitational-Wave Observatory (LIGO)
Scientific Collaboration and Virgo Collaboration
opens the window to study strong field gravitational physics
and test alternative theories of gravity \cite{Abbott:2016blz,Abbott:2016nmj,Abbott:2017vtc,Abbott:2017gyy,Abbott:2017oio,TheLIGOScientific:2017qsa}.
In particular, the recent detection of the GW170817 from the merger of a binary neutron star \cite{TheLIGOScientific:2017qsa}
and the electromagnetic counterparts starts a new era of multi-messenger GW astronomy.
A gamma ray burst GRB170817A was observed $1.74\pm 0.05$s later by Fermi Gamma-Ray Burst Monitor \cite{Goldstein:2017mmi}
and the International Gamma-Ray Astrophysics Laboratory \cite{Savchenko:2017ffs}.
If we assume that the peak of the GW signal and the first photons were emitted simultaneously, and the $1.74$s time difference is
caused by the faster speed of GWs, then we get an upper bound on the speed of GWs $c_{gw}/c-1\le 7\times 10^{-16}$ \cite{Monitor:2017mdv}.
If we assume that the GRB signal was emitted 10s after the GW signal, then we get a lower bound $c_{gw}/c-1>-3\times 10^{-15}$ \cite{Monitor:2017mdv}.
The precise measurement of the propagation speed of GWs is a very powerful tool to test alternative theories
of gravity \cite{Mirshekari:2011yq,Jimenez:2015bwa,Chesler:2017khz,Baker:2017hug,Creminelli:2017sry,Sakstein:2017xjx,Ezquiaga:2017ekz,Green:2017qcv,Nishizawa:2017nef,Arai:2017hxj,Battye:2018ssx}

Recently there is a lot of activity studying  scalar-tensor theories \cite{Fujii:2003pa} and one of
them is the gravitational theory which is the result of  the Horndeski Lagrangian
\cite{Horndeski:1974wa}. Horndeski theories because  they
lead to second-order field equations can be technically simple, and they prove
consistent  without ghost instabilities \cite{Ostrogradsky:1850fid}.
In Horndeski theory the derivative self-couplings of the scalar field screen the
deviations from GR at high gradient regions (small scales or high
densities) through the Vainshtein mechanism \cite{Vainshtein:1972sx},
thus satisfying solar system and early universe constraints
\cite{Nicolis:2008in,Deffayet:2009wt,Chow:2009fm,DeFelice:2011th,Babichev:2011kq,Chakraborty:2012sd,Lambiase:2015yia,Bhattacharya:2016naa}.

A subclass of Horndeski theories includes the coupling of  the scalar field to Einstein tensor. This term
introduces a new mass scale in the theory which on short distances allows
to find  black hole solutions \cite{Kolyvaris:2011fk,Rinaldi:2012vy,Kolyvaris:2013zfa,Babichev:2013cya,Charmousis:2014zaa},
while a black hole can be formed if one considers the gravitational collapse of a scalar field coupled to the Einstein tensor \cite{Koutsoumbas:2015ekk}.
On large distances  the presence of the derivative coupling acts as a friction term in the inflationary period of the cosmological
evolution  \cite{Amendola:1993uh,Sushkov:2009hk,Germani:2010hd,Saridakis:2010mf,Huang:2014awa,Yang:2015pga}. Also, the preheating  period  at the end of
inflation was studied, and it was found that  there is a suppression of heavy particle production  as the derivative coupling is increased. This was attributed to the fast decrease of  kinetic
energy of the scalar field because of  its  wild oscillations \cite{Koutsoumbas:2013boa}. A holographic application was performed in \cite{Kuang:2016edj} where it was shown that the change of the kinetic energy of the scalar field coupled  to Einstein tensor allowed to holographically simulate the effects of a high concentration of impurities in a real material.
The above discussion indicates that the coupling of the scalar field to Einstein tensor alters the kinematical properties of the scalar field.

Assuming that the scalar field coupled to Einstein tensor plays the role of dark energy and drives the late cosmological expansion it was  found \cite{Germani:2010gm,Germani:2011ua} that
the propagation speed of the tensor perturbations around the cosmological background with
Friedmann-Robertson-Walker (FRW) metric is different from the speed of light $c$, so the measurement of
the current speed of GWs can be used to test the applicability of this and Horndeski theories
to explain the late time accelerated cosmological expansion \cite{Lombriser:2015sxa,Lombriser:2016yzn,Bettoni:2016mij,Baker:2017hug,Creminelli:2017sry,Sakstein:2017xjx,Ezquiaga:2017ekz}.
Due to small deviation (on the order of $10^{-15}$) of the current speed of GWs from the speed of light,
in Refs. \cite{Baker:2017hug,Creminelli:2017sry,Ezquiaga:2017ekz}
it was argued that dark energy models that predict $c_{gw}\neq c$ at late cosmological times are ruled out,
and in Horndeski theory the only viable nonminimal coupling to gravity has the conformal form $f(\phi)R$.
However, with the help of derivative conformal or disformal transformations,
some Horndeski and beyond Horndeski theories can survive the speed constraint \cite{Ezquiaga:2017ekz,Babichev:2017lmw,Babichev:2018uiw,Babichev:2013cya}.

In this work, we will perform a detailed analysis on the effect of the latest observational results on the current speed of GWs $c_{gw}$ to
the Horndeski theories with the non-minimally derivative coupling. From the early cosmological evolution we know that the derivative coupling
of the scalar field to Einstein tensor alters the kinetic energy of the scalar field \cite{Sushkov:2009hk}
influencing in this way the dynamical evolution of the Universe giving an upper bound to the mass scale coupling.
For the late cosmological evolution we will perturb the FRW
metric under tensor perturbations and we will show that a subclass of Horndeski theory
consisting of the usual kinetic term and the coupling of the scalar field to Einstein tensor
is still viable provided that the mass scale $M$ introduced in the non-minimally derivative
coupling is highly constrained from the recent results on the speed of GWs.
The result also shows that no tuning on the parameter is needed to satisfy the stringent constraints on the speed of GWs.
For comparison we also discuss bounds on the Gauss-Bonnet coupling from the
observational bounds on $c_{gw}$.

The paper is organized as follows. In Sect. 2 we discuss the Horndeski theory,
and studying the speed of tensor perturbations  we obtain bounds on the mass scale
introduced by the presence of the derivative coupling of the scalar field to Einstein tensor.
In Sect. 3 we discuss the bounds on coupling $\alpha$ of the
Gauss-Bonnet theory coupled to a scalar field. Finally, in Sect. 4 are our conclusions.

\section{The effect of the speed of gravitational waves in the Horndeski theory}

In this section we will briefly review the Horndeski theory, we will discuss the speed of GWs in this theory
and assuming that the scalar field present in the Horndeski Lagrangian plays the role of dark energy we will find a lower bound on the mass scale introduced in derivative coupling of the scalar field to Einstein tensor.

The action of the Horndeski theory is given by \cite{Horndeski:1974wa},
\begin{equation}
\label{acth}
S=\int d^4x\sqrt{-g}(L_2+L_3+L_4+L_5)~,
\end{equation}
where
\begin{gather*}
L_2=K(\phi,X)~,\quad L_3=-G_3(\phi,X)\Box \phi~, \\ L_4=G_4(\phi,X)R+G_{4,X}\left[(\Box\phi)^2-(\nabla_\mu\nabla_\nu\phi)(\nabla^\mu\nabla^\nu\phi)\right]~, \\
L_5=G_5(\phi,X)G_{\mu\nu}\nabla^\mu\nabla^\nu\phi-\frac{1}{6}G_{5,X}[(\Box\phi)^3\\
-3(\Box\phi)(\nabla_\mu\nabla_\nu\phi)(\nabla^\mu\nabla^\nu\phi)+2(\nabla^\mu\nabla_\alpha\phi)(\nabla^\alpha\nabla_\beta\phi)(\nabla^\beta\nabla_\mu\phi)]~,
\end{gather*}
with $X=-\nabla_\mu\phi\nabla^\mu\phi/2$, $\Box\phi=\nabla_\mu\nabla^\mu\phi$,
the functions $K$, $G_3$, $G_4$ and $G_5$ are arbitrary functions of $\phi$ and $X$, and $G_{j,X}(\phi,X)=\partial G_j(\phi,X)/\partial X$ with $j=4,5$.

This action is the most general one for scalar-tensor theory with at most second-order field equations. If we take $K=G_3=G_5=0$ and $G_4=M_{\text{Pl}}/2$,
then we obtain Einstein's general relativity. If we take $G_3=G_5=0$, $K=X-V(\phi)$, and $G_4=f(\phi)$, then we get scalar-tensor $f(\phi) R$ theories.
If we take $G_4=M_{\text{Pl}}^2/2+X/(2M^2)$ or $G_4=M_{\text{Pl}}^2/2$ and $G_5=-\phi/(2M^2)$,
then we get the non-minimally derivative coupling $G_{\mu\nu}\nabla^\mu\phi\nabla^\nu\phi/(2M^2)$ with  the mass scale $M$  \cite{Kobayashi:2011nu}.

The stability of the Horndeski theory in the FRW background was studied in \cite{DeFelice:2011bh}. General conditions on the functions appearing in the Horndeski Lagrangian were given for the theory to  be ghost free and stable under tensor perturbations. While in the flat background, the propagation speed of tensor perturbations is the same as the speed of light \cite{Hou:2017bqj}, in the cosmological FRW background  the propagation speed of tensor perturbations in the Horndeski theory was found to be \cite{Kobayashi:2011nu}
\begin{equation}
\label{hornct1}
c_{gw}^2=\frac{G_4-X\left(\ddot{\phi}G_{5,X}+G_{5,\phi}\right)}{G_4-2 XG_{4,X}-X\left(H\dot{\phi}G_{5,X}-G_{5,\phi}\right)}~.
\end{equation}

We are interested in the propagation speed of tensor perturbations of the subclass of the Horndeski theory that consists of the usual kinetic scalar field term and the coupling of the scalar field to Einstein tensor given by the action
\begin{equation}\label{action}
S=\int d^4x\sqrt{-g}\left\{\frac{M_{\text{Pl}}^2}{2}R-\frac{1}{2}\left[g_{\mu\nu}-\frac{1}{M^2} G_{\mu\nu}\right]\nabla^\mu \phi \nabla^\nu \phi-V(\phi)\right\}~.
\end{equation}
Perturbing the FRW metric as
\begin{equation}\label{tensor:per}
ds^2=-d t^2+a^2\left(t\right)\left(\delta_{ij}+h_{ij}\right)dx^i dx^j,
\end{equation}
and expanding the action  \eqref{action} to the second order of the tensor perturbations $h_{ij}$, we obtain the quadratic action \cite{Yang:2015pga,Germani:2011ua}
\begin{equation}
\label{action2}
S=\frac{ M_{\text{Pl}}^2}{ 8}\int d^3xdt a^3 \left[\left(1-\Gamma\right)\dot{h}_{ij}^2-\frac{1}{a^2}\left(1+\Gamma\right)\left(\partial_k h_{ij}\right)^2\right],
\end{equation}
where $\Gamma= \dot{\phi}^2/(2M^2 M_{\text{Pl}}^2)$. From the action \eqref{action2} we derive the equation of  motion for GWs
\begin{equation}
\label{eq1}
  \ddot{h}_{ij}+3H\dot{h}_{ij}-\frac{\dot{\Gamma}}{1-\Gamma}\dot{h}_{ij}-\frac{1+\Gamma}{1-\Gamma}\frac{\nabla^2h_{ij}}{a^2}=0.
\end{equation}
Under the transverse-traceless gauge, the Fourier components of tensor perturbations $h_{ij}(\vec{x},t)$ is
\begin{equation}\label{FT}
h_{ij}(\vec{x},t)=\int d^3 k \left[h_k^+\left(t\right)\epsilon^+_{ij}+h_k^{\times}\left(t\right)\epsilon^{\times}_{ij}\right]
\exp\left(i \vec{k}\cdot \vec{x}\right),
\end{equation}
where $k^i\epsilon_{ij}^s=\epsilon_{ii}^s=0$, $\epsilon_{ij}^s\epsilon_{ij}^{s'}=2\delta_{ss'}$,
and the superscript ``$s$" stands for the ``$+$" or ``$\times$" polarizations. Substituting Eq. \eqref{FT} into Eq. \eqref{eq1},
we get
\begin{equation}
\label{eq2}
  \ddot{h}_k^s+3H\dot{h}_k^s\left[1-\frac{\dot{\Gamma}}{3H\left(1-\Gamma\right)}\right]+\frac{c^2_{gw} k^2}{a^2}h_k^s=0.
\end{equation}
The propagation speed for both polarization states is
\begin{equation}
\label{cgeq1}
c_{gw}^2=\frac{1+\Gamma}{1-\Gamma}.
\end{equation}
This result can also be obtained from the general formula of the Horndeski theory given in equation  \eqref{hornct1} choosing  $G_4=M_{\text{Pl}}^2/2+X/(2M^2)$ or $G_4=M_{\text{Pl}}^2/2$ and $G_5=-\phi/(2M^2)$.
For the Horndeski theory, it was argued that the precise measurement
on the speed of GWs $c_{gw}=1$ requires $G_{4,X}=G_{5,\phi}=G_{5,X}=0$,
and only the conformal coupling $f(\phi)R$ is allowed \cite{Baker:2017hug,Creminelli:2017sry,Ezquiaga:2017ekz}.
At late times, since the effect of the non-minimally derivative
coupling $G_{\mu\nu}\nabla^\mu\phi\nabla^\nu\phi\sim H^2 \dot\phi^2/M^2$ is
negligible compared with the canonical kinetic term $\dot\phi^2$
due to the decrease of the Hubble parameter $H$ as the Universe expands, the speed given by Eq. \eqref{cgeq1} can be close to 1.
Instead of requiring that $G_{4,X}=G_{5,\phi}=G_{5,X}=0$, we show that a large mass range for the coupling $M$
is allowed to satisfy the stringent constraint on the speed of GWs with negligible but nonzero deviation.

Using the the upper bound on the speed of of GWs \cite{Monitor:2017mdv}
\begin{equation}
\label{speed:condition}
\frac{c_{gw}}{c}-1\leq 7 \times 10^{-16},
\end{equation}
we obtain,
\begin{equation}
\label{gbld1}
0< \Gamma \leq 7\times 10^{-16}.
\end{equation}
This constraint is much less stringent than the classical constraint $\Gamma\leq 2/3\times10^{-20}$
derived from the constraint on the Parameterized Post-Newtonian (PPN) parameter $\alpha_3=6\Gamma<4\times 10^{-20}$ \cite{Zhu:2015lry}.

Using the constraint (\ref{gbld1}) we will obtain a lower bound on the mass scale $M$ of the derivative coupling.
If we take the scalar field as dark energy and use the observational constraint
$1+w=\dot\phi^2/\rho_\phi=\dot\phi^2/(3M^2_{\text{Pl}}H^2_0\Omega_\phi)\sim 0.2$ and $\Omega_m=0.3$,
then we get the contribution of the canonical kinetic energy as
 \begin{equation}
 \label{debldeq}
   \frac{(1+w)(1-\Omega_m)}{2}=\frac{\dot\phi^2}{6M^2_{\text{Pl}}H^2_0}=\frac{\Gamma}{3}\frac{M^2}{H^2_0}\sim 0.07.
 \end{equation}
Combining eqs. \eqref{gbld1} and \eqref{debldeq}, we get the constraint on the coupling constant $M$
\begin{equation}
\label{gwbldeq1}
\frac {H_0^2}{M^2} \lesssim  3.3 \times 10^{-15}.
\end{equation}
For the theory with the coupling of a scalar field to Einstein tensor, the PPN parameters are \cite{Zhu:2015lry}
\begin{equation}
\label{ppneq1}
\begin{split}
\beta=1+6\Gamma,\ \ \gamma=1+3\Gamma, \ \  \alpha_1=12\Gamma,\ \ \alpha_2=3\Gamma,\\
\alpha_3=6\Gamma,\ \ \zeta_2=15\Gamma,\ \  \zeta_3=3\Gamma,\ \ \xi=\zeta_1=\zeta_4=0.
\end{split}
\end{equation}
The stringent constraint coming from $\alpha_3=6\Gamma<4\times 10^{-20}$ \cite{Bell:1996ir,Stairs:2005hu}, substituting this result into
eq. \eqref{debldeq}, we get
\begin{equation}
\frac{H_0^2}{M^2} \lesssim  3.2\times 10^{-19}.
\end{equation}
This constraint is much stronger than eq. \eqref{gwbldeq1}.
If we use the constraint \eqref{gwbldeq1}, we get the lower bound on the coupling $M$ as $M \gtrsim 2\times 10^{-35}$GeV.
In the New Higgs inflation \cite{Germani:2010gm}, the non-minimally derivative coupling enhances the friction of the expansion
and the high friction limit requires $M\ll 10^{15}$ Gev \cite{Yang:2015pga} while in \cite{Dalianis:2016wpu}  limits on $M$ are also discussed   during the reheating period. Therefore, the
mass scale introduced in the  derivative coupling is $10^{15}\text{GeV}\gg M \gtrsim 2\times 10^{-35}$GeV \footnote{This lower bound of the mass $M$ allows the sound speed squared of the tensor perturbations to reach the observational bounds of the GWs in the model discussed in \cite{Koutsoumbas:2017fxp} making in this way the model of unification of dark matter with dark energy in Horndeski theory viable. }.

This result is interesting and it shows that the coupling of the scalar field to Einstein tensor has a complete different behaviour compared to a scalar field minimally coupled to gravity. While the kinetic energy of a minimally coupled scalar field practically does not understand the cosmological evolution, the kinetic energy of scalar field coupled to Einstein tensor changes as the Universe expands. At the inflationary epoch it can drive inflation with steep potentials while as the Universe expands its contribution to the cosmological evolution is less important and at the late cosmological epoch is negligible, so GWs propagate at the speed of light at late times.

\section{The effect of the speed of gravitational waves in the  Gauss-Bonnet theory}

Another non-trivial extension of GR which gives second order differential equations is the Lovelock theory \cite{Lovelock:1971yv},
which apart from the Einstein-Hilbert term also includes higher order curvature terms. The simplest case is the Gauss-Bonnet theory which is a second order Lovelock theory which however in four dimensions is a topological invariance. If however it is coupled to a scalar field then a scalar-tensor theory is generated from the action \cite{Rizos:1993rt},
\begin{equation}
S=\int\sqrt{-g}d^4 x\left[\frac{M_{\text{Pl}}^2}{2}R+X-V(\phi)+\frac{f\left(\phi\right)}{8}R_{GB}\right]~,
\end{equation}
where
\begin{equation}
R_{\text{GB}}=R^2-4R_{\mu\nu}R^{\mu\nu}+R_{\mu\nu\alpha\beta}R^{\mu\nu\alpha\beta}~,
\end{equation}
and we have also included a scalar potential $V(\phi)$.

It is interesting to notice that the Gauss-Bonnet theory coupled to a scalar field in four dimensions
can be generated from the general Horndeski action (\ref{acth}) \cite{Kobayashi:2011nu,DeFelice:2011uc} making the  following identifications  of the functions involved
\begin{gather}
  K\left(\phi,X\right)=X-V(\phi)+f''''\left(\phi\right)X^2\left(3-\ln X\right)~,\label{funct1}\\
  G_3\left(\phi,X\right)=\frac{f'''\left(\phi\right)}{2} X \left(7-3\ln X\right)~,\\
  G_4\left(\phi,X\right)=\frac{M_{\text{Pl}}^2}{2}+\frac{f''\left(\phi\right)}{2}X\left(2-\ln X\right)~,\\
  G_5\left(\phi,X\right)=-\frac{f'(\phi)}{2} \ln X~. \label{funct2}
  \end{gather}
Then we can use the general formula for the propagation of gravitational waves equation  \eqref{hornct1} for the functions (\ref{funct1})-(\ref{funct2}) and  we get
\begin{equation}
c_{gw}^2=\frac{M_{\text{Pl}}^2+2Xf''\left(\phi\right)+\ddot{\phi}f'\left(\phi\right)}{M_{\text{Pl}}^2+H\dot{\phi}f'\left(\phi\right)}~. \label{grav1}
\end{equation}

The role that the Gauss-Bonnet tern coupled to a scalar field  plays in the late cosmological
evolution has been extensively studied \cite{Nojiri:2005vv,Koivisto:2006xf,Koivisto:2006ai,Calcagni:2006ye,
Carter:2005fu,Copeland:2006wr,Leith:2007bu,Nozari:2017rta,Guo:2009uk,Koh:2014bka,Kanti:2015dra,Koh:2016abf,Heydari-Fard:2016nlj}.
In this work we will use a specific model discussed in  \cite{Heydari-Fard:2016nlj}. In this model by choosing the scalar potential as
\begin{equation}
\label{gbdemod1}
V(\phi)=\left[-\frac{\beta}{2}+\frac{1}{9}(3-8\alpha)\right]g(\phi)-\rho_0e^{-\phi/\sqrt{\beta}}~,
\end{equation}
where $\alpha$ and $\beta$ are model parameters.
An exact solution of the gravitational field equations $\phi=3\sqrt{\beta}\ln a$ was found with the coupling function of the scalar field to be
\begin{equation}
\label{gbdemod2}
f(\phi)=\alpha \int \frac{8(3-4\alpha)d\phi}{3\sqrt{\beta} g(\phi)}~,
\end{equation}
where
\begin{equation}
\label{gbdemod3}
g(\phi)=\frac{(3-4\alpha)A}{3}\exp\left(\frac{(8\alpha+27\beta)\phi}{3\sqrt{\beta}(4\alpha-3)}\right)
+\frac{9(4\alpha-3)\rho_0}{20\alpha+27\beta-9}e^{-\phi/\sqrt{\beta}}~,
\end{equation}
and $A$ is an integration constant, $\rho_0$ is the present value of the energy density for matter. Note that if the coupling strength $\alpha=0$ we do not have the Gauss-Bonnet coupling. From the solution, we get the current value of the ratio of the energy densities between dark energy
and matter,
\begin{equation}
\label{rhodedm1}
\frac{\Omega_{de}}{\Omega_m}=\frac{A (20 \alpha +27 \beta -9)-3 \rho _0 (20 \alpha +27 \beta )}{3 \rho _0 (20 \alpha +27 \beta -9)},
\end{equation}
and the current equation of state parameter for dark energy
\begin{equation}
\label{wdeeq1}
w_{de}=-\frac{A (20 \alpha +27 \beta -9)^2}{3 (4 \alpha -3) \left(c (20 \alpha +27 \beta -9)-3 \rho _0 (20 \alpha +27 \beta )\right)}.
\end{equation}
If we take the current value of the ratio $\Omega_{de}/\Omega_m$ to be $7/3$ \cite{Ade:2015xua}, then we obtain the integration constant $A$,
\begin{equation}
\label{gbdemod4}
A=\frac{(200\alpha+270\beta-63)\rho_0}{20\alpha+27\beta-9}~,
\end{equation}
and the  current equation of state parameter for dark energy
\begin{equation}
\label{gbdemod5}
w_{de}=-\frac{63-200\alpha-270\beta}{63-84\alpha}~.
\end{equation}

\begin{figure}[htp]
\centerline{\includegraphics[width=0.4\textwidth]{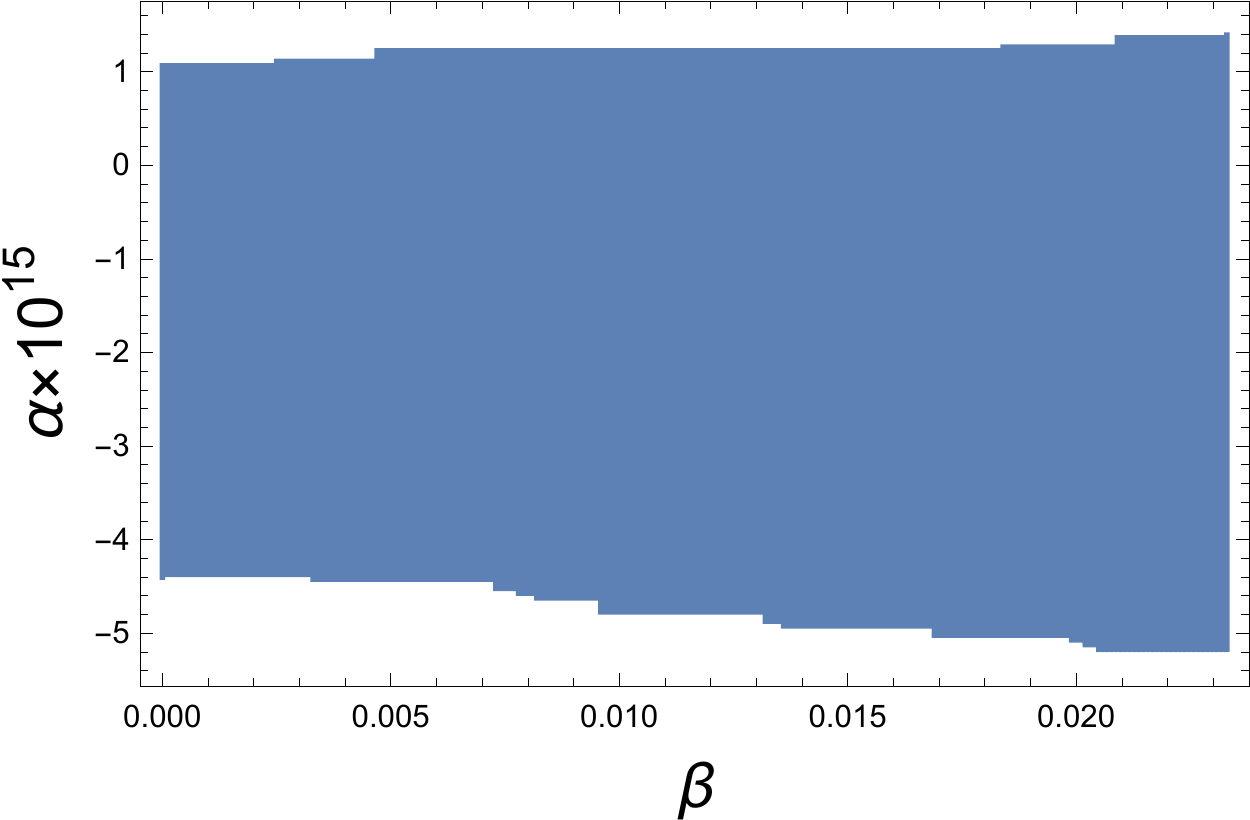}}
\caption{The constraint on the coupling constant $\alpha$ and the model parameter $\beta$.}
\label{fig1}
\end{figure}

Using eq. \eqref{grav1} the speed of GWs  is
\begin{equation}
\label{gbdemod6}
c^2_{gw}=1-\frac{320 \alpha ^2+6 \alpha  (90 \beta -11)}{45+160 \alpha ^2-180 \alpha}~.
\end{equation}
Using the bound on the speed of GWs \cite{Monitor:2017mdv}
\begin{equation}
\label{speed:condition1}
-3\times 10^{-15}\le \frac{c_{gw}}{c}-1\leq 7 \times 10^{-16},
\end{equation}
and the constraint $-1.1<w_{de}<-0.9$ \cite{Ade:2015xua}, we get the range on the model parameters $\alpha$ and $\beta$ as shown in Fig. \ref{fig1}.

The results in Fig. \ref{fig1} show that for a range of values of the parameter $\beta$, the coupling strength of the Gauss-Bonnet term is in the order of $10^{-15}$.
For the inflationary model with $V(\phi)\sim \phi^2/2$ and $f(\phi)=-8\alpha\phi^2/2$, the absolute value of the coupling strength $\alpha$ is constrained
to be less than the order of $0.01$ \cite{Nozari:2017rta}. For the power-law inflation with both exponential coupling and potential,
the coupling strength is constrained to be $-1\times 10^{-4}<\alpha<4\times 10^{-4}$ \cite{Guo:2009uk}. Therefore, the constraint from the
speed of GWs is much stronger.

\section{Conclusions}

The first measurement of the speed of GWs by GW170817 and GRB170817A bounds the deviation of
the speed of GWs from the speed of light to be no more than one part in $10^{15}$, so it provides
the evidence that $c_{gw}=c$. Using these observational result we  can test alternative theories of gravity for their validity
to describe the cosmological evolution at late times.

We used these bounds on $c_{gw}$ to constrain first a subclass of the Horndeski theory
in which a scalar field except its minimal coupling is also coupled to Einstein tensor. Assuming that the scalar field plays the role of dark energy we found a lower bound on the mass scale introduced by this coupling and combining the constraints from inflation the energy scale of the derivative coupling is bounded to be $10^{15}\text{GeV}\gg M \gtrsim 2\times 10^{-35}$GeV.
This result requires no fine-tuning and shows that it is possible to get $c_{gw}\approx c$
from the terms with $G_{4,X}\neq 0$ and $G_{5,\phi}\neq 0$ if their effects are negligible at late times.

We also studied the Gauss-Bonnet theory in four dimensions coupled to a scalar field.  The coupling
of the Gauss-Bonnet term to scalar field not only gives successful inflation, but also provides late time cosmic acceleration.
Using a particular model with a specific form of the coupling function $f(\phi)$ which allows an exact solution of the gravitational equations, we found that the bounds on $c_{gw}$ constrains the Gauss-Bonnet coupling strength to be $\alpha\lesssim 10^{-15}$, a constraint much stronger than the coupling strength  $-1\times 10^{-4}<\alpha<4\times 10^{-4}$
resulted from models with  power-law inflation.

\begin{acknowledgement}
We thank K. Ntrekis,  S. Tsujikawa, I. Dalianis, J. Sakstein, L. Lombriser, M. Zumalacarregui
and D. Shantanu for useful discussions.
E.P acknowledges the hospitality of School of Physics of Huazhong University of Science and Technology
where part of this work was carried out.
This research was supported in part by the Major Program of the National Natural Science Foundation of China under Grant No. 11690021
and the National Natural Science Foundation of China under Grant Nos. 11875136 and 11475065.
\end{acknowledgement}

%\bibliographystyle{spphys}
%\bibliography{../../book/cosmologyref}

\end{document}